\documentclass[pre,twocolumn,aps,showpacs]{revtex4}

\newcommand{\beq}{\begin{equation}}
\newcommand{\eeq}{\end{equation}}
\newcommand{\bqn}{\begin{eqnarray}}
\newcommand{\eqn}{\end{eqnarray}}
\newcommand{\bqns}{\begin{eqnarray*}}
\newcommand{\eqns}{\end{eqnarray*}}
\newcommand{\bary}{\begin{array}}
\newcommand{\eary}{\end{array}}
\newcommand{\non}{\nonumber}

\usepackage{graphicx}%
\usepackage{amsmath}

\begin{document}
\title{Note on the positiveness of magnetic energy stored in the Wire-SRR metamaterial}
\author{Pi-Gang Luan}
\address{Wave Engineering Laboratory, Department of Optics and
Photonics, National Central University, Jhungli 320, Taiwan}

\begin{abstract}%
Recently, a new energy density formula for electromagnetic waves
in the metamaterial consisting of arrays of conducting wires and
split-ring-resonators (SRR) has been derived [Phys. Rev. E 80,
046601 (2009)]. According to that formula, the positiveness is
obvious for the electric part of the energy density but not clear
for the magnetic part. In this paper, I show that the magnetic
energy density is also positively definite.
\end{abstract}
\pacs{41.20.Jb, 03.50.De, 78.20.Ci} \maketitle
In a recent paper, Luan derived a new expression for energy
density of electromagnetic waves in the metamaterial medium
consisting of wires and split-ring-resonators (SRR) \cite{Luan}.
The motivation of that paper was to resolve the apparent
contradictions between the results given by the equivalent circuit
(EC) and electrodynamic (ED) approaches \cite{Tretyakov,
Boardman}. Luan pointed out a calculation error made in
\cite{Tretyakov}, which can be corrected so that EC approach can
really give the correct formula. On the other hand, usually the ED
approach \cite{Boardman} does not give unique result, and the
analysis in \cite{Luan} revealed that the expression of the energy
density is in fact determined by the power loss we choose. For a
lossy medium, Luan suggested to find the correct expression of
power loss first, then the correct energy density formula will
emerge automatically. The equivalence between EC and ED approaches
was then confirmed by reproducing the EC energy formula after time
averaging the ED formula.

The positiveness of the electric energy density can be easily
observed from the following expressions \beq
W_e=\frac{\epsilon_0{\bf
E}^2}{2}+\frac{1}{2\omega^2_p\epsilon_0}\left(\frac{\partial{\bf
P}}{\partial t}\right)^2 \label{We},\eeq \beq \left\langle
W_e\right\rangle =\frac{\epsilon_0|{\bf
E}|^2}{4}\left(1+\frac{\omega^2_p}{\omega^2+\nu^2}\right). \eeq
Although the positiveness of magnetic energy density in the
time-domain formula is obvious (note that $0<F<1$ is assumed) \bqn
W_b&=&\frac{\mu_0(1-F)}{2}{\bf H}^2+\frac{\mu_0}{2F}\left({\bf
M}+F{\bf H}\right)^2
\\&&+\frac{\mu_0}{2\omega^2_0F}\left(\frac{\partial {\bf M}}{\partial t}
+F\frac{\partial {\bf H}}{\partial t}+\gamma{\bf
M}\right)^2,\label{Wb}\eqn this property is not clear in the time
averaged formula \beq \left\langle
W_b\right\rangle=\frac{\mu_0|{\bf
H}|^2}{4}\left[1+F\frac{\omega^2\left(3\omega^2_0-\omega^2\right)}
{\left(\omega^2_0-\omega^2\right)^2+\omega^2\gamma^2}\right].
\label{wbharmonic}\eeq

When $\omega>\sqrt{3}\omega_0$, the above formula gives a $\langle
W_b\rangle$ smaller than the vacuum value $\mu_0|{\bf H}|^2/4$.
Will this expression always give a positive  $\langle W_b\rangle$
? If we add the two terms in the square bracket directly, we get
\beq \frac{4\langle W_b\rangle}{\mu_0|{\bf
H}|^2}=\frac{(1-F)\omega^4+(3F-2)\omega^2_0\omega^2+\omega^4_0+\omega^2\gamma^2}
{(\omega^2_0-\omega^2)^2+\omega^2\gamma^2},\eeq which is still
hard to judge its sign. However, if we consider the identity \bqn
&&(\omega^2_0-\omega^2)^2
+\omega^2\gamma^2+F\omega^2(3\omega^2_0-\omega^2)\non\\
&&=(1-F)(\omega^2-\omega^2_0)^2+F\omega^2_0(\omega^2+\omega^2_0)+\omega^2\gamma^2
\eqn the positiveness of the magnetic energy becomes very obvious
(remember that both $F$ and $1-F$ are positive). According to this
analysis, both the electric and magnetic energy densities in the
energy density formula of \cite{Luan} are positively definite.

\section*{ACKNOWLEDGEMENT} The author gratefully acknowledge
financial support from National Science Council (Grant No. NSC
98-2112-M-008-014-MY3) of Republic of China, Taiwan.


\end{document}